\documentclass[12pt,letterpaper]{article}
\usepackage{jheppub}
\usepackage{graphicx} 
\usepackage{amsmath}
\usepackage[]{amssymb}

\usepackage{epsfig}
\newcommand{\be}{\begin{equation}}
\newcommand{\ee}{\end{equation}}
\newcommand{\beq}{\begin{equation}}
\newcommand{\eeq}{\end{equation}}
\newcommand{\bea}{\begin{eqnarray}}
\newcommand{\eea}{\end{eqnarray}}

\def\be{\begin{equation}}
\def\ee{\end{equation}}
\def\ba{\begin{eqnarray}}
\def\ea{\end{eqnarray}}

\DeclareMathOperator{\Tr}{Tr}

\title{
Entropy on a null surface for interacting quantum field theories and
the Bousso bound}

\author[a,b]{Raphael Bousso,}
\author[c,d]{Horacio Casini,}
\author[a,b]{Zachary Fisher,}
\author[d]{and Juan Maldacena}
\affiliation[a]{Center for Theoretical Physics and Department of Physics,\\
 University of California, Berkeley, CA 94720, U.S.A.}
\affiliation[b]{Lawrence Berkeley National Laboratory, Berkeley, CA 94720,
  U.S.A.}
\affiliation[c]{Centro At\'omico Bariloche, 8400, Bariloche, R\'{\i}o Negro, Argentina}
\affiliation[d]{Institute for Advanced Study, Princeton, NJ 08540, USA}

\abstract{ 
We study the vacuum-subtracted von Neumann entropy of a segment on a null plane. We argue that for 
interacting quantum field theories in more than two dimensions,  this entropy has a simple expression 
in terms of the expectation value of the null components of the stress tensor on the null interval. 
More explicitly $\Delta S = 2\pi  \int d^{d-2}y \int_0^1 dx^+\, g(x^+)\, \langle T_{++}\rangle$, where  
$g(x^+)$ is a theory-dependent function. This function is constrained by general properties of quantum relative entropy. 
These constraints are enough to extend our recent free field proof of the quantum Bousso bound to the interacting case. 

This unusual expression for the entropy as the  expectation value of an operator implies that the entropy is 
equal to the modular Hamiltonian, $\Delta S = \langle \Delta K \rangle  $, where $K$ is the operator in the right hand side. 
We explain how this equality is compatible with a non-zero value for $\Delta S$. 
Finally, we also compute explicitly the function $g(x^+)$ for theories that have a gravity dual. 
}

\begin{document}
\maketitle

\section{Introduction}

In a recent paper~\cite{BouCas14}, we proved the Bousso bound, or covariant entropy bound~\cite{CEB1}, 
\begin{equation}
\Delta S \leq \frac{A-A'}{4G\hbar}~,
\end{equation}
for light-sheets with initial area $A$ and final area $A'$~\cite{FMW}.\footnote{The search for a holographic entropy bound in general spacetimes was inspired by~\cite{Tho93,Sus95,FisSus98}; see \cite{RMP} for a review.} The proof applies to free fields, in the limit where gravitational back-reaction is small, $G\hbar\to 0$, 
that the change in the area is of first order in $G$. 

Though this regime is limited, the proof is interesting. No assumption is needed about the relation between the entropy and energy of quantum states, beyond what quantum field theory already supplies. Conversely, this suggests that 
 quantum gravity may determine some properties of local field theory in the weak gravity limit.

In the present paper, we will generalize our proof to interacting theories. We will continue to work in the weakly gravitating regime. In the course of this analysis, we will establish a number of interesting properties of the entropy and modular energy on finite planar light-sheets, for general interacting theories.

In the free case, we defined the entropy as the difference of two von Neumann entropies~\cite{Cas08,MarMin04}. The relevant states are the reduced density operators of an arbitrary quantum state and the vacuum, both obtained by tracing over the exterior of the light-sheet. Following Wall~\cite{Wal11}, we were able to work directly on the light-sheet.

Let us recall the structure of the proof in the free case. A very general result, the positivity of the relative entropy~\cite{Lin73}, implies that $\Delta S\leq \Delta K$, where $\Delta K$ is the vacuum-subtracted expectation value of the modular Hamiltonian operator\footnote{For any state $\rho_1$, the modular energy is $\Delta K \equiv \Tr\, (K\rho_1) - \Tr\, (K\rho_0)$.  The modular Hamiltonian $K$ is the logarithm of the vacuum density matrix $K=-\log \rho_0$. $K$ is defined up to an additive constant, which can be fixed by requiring that the vacuum expectation value of $K$ is zero, such that $\Delta K=\langle K\rangle$.
Similarly, $\Delta S = - \Tr[ \rho_1 \log \rho_1] + \Tr[ \rho_0 \log \rho_0] $ is the difference between the entropy for the state $\rho_1$ under consideration and
the vacuum $\rho_0$. 
}~\cite{Cas08}. For free theories, the modular energy is found to be given by an integral over the stress tensor, 
\begin{equation}
% \Delta K = \frac{2\pi}{\hbar} \int d^{d-2}y \int_0^{\Delta x^+} d\hat{x}^+\, \Delta x^+\, g(\hat{x}^+/\Delta x^+)\, T_{++}(\hat{x}^+,y)~.
\Delta K = \frac{2\pi}{\hbar} \int d^{d-2}y \int_0^1 dx^+\, g(x^+)\, T_{++}(x^+,y)~.
\label{eq-dkgen}
\end{equation}
Here $x^+$ is an affine parameter along the null generators, which can be scaled so that the null interval has unit length. The function $g$ is given by
\begin{equation}
g(x^+) = x^+ (1-x^+)~.
\label{eq-gfree}
\end{equation}
(For $d=2$, $g$ takes this form also in the interacting case; but as we shall see, in higher dimensions it will not.)

By Einstein's equation, the area difference $\Delta A = A-A'$ is also given by a local integral over the stress tensor, plus a term that depends on the initial expansion of the light-rays.  The latter must be chosen so that the expansion remains nonpositive everywhere on the null interval. This is the ``non-expansion condition'' that determines whether a null hypersurface is a light-sheet. Eqs.~(\ref{eq-dkgen}) and (\ref{eq-gfree}), combined with Einstein's equation and the nonexpansion condition, imply that $\Delta K\leq\Delta A/4G\hbar$.

To generalize this proof to interacting theories, a number of difficulties must be addressed. Wall's results do not apply, so the entropy and modular Hamiltonian cannot be defined directly on the light-sheet. Instead, we must consider spatial regions that approach the light-sheet. The positivity of the relative entropy, $\Delta K - \Delta S \ge 0$, holds for every spatial region \cite{Cas08}, so it could still be invoked. But it is no longer useful: for spatial regions, $\Delta K$ is highly nonlocal, and we are unable to compute it before taking the null limit.

Instead, we benefit from a new simplification, which happens to arise precisely in the case to which our previous proof did not apply: for interacting theories in $d>2$.\footnote{Our original proof applies to theories for which the algebra of observables is nontrivial and factorizes between null generators. This includes free theories but also interacting theories in $d=2$~\cite{Wal11}. For $d=2$, the area  is the expectation value of the dilaton-like field $\Phi$ that appears in the action as 
$ { 1 \over 16 \pi G } \int d^2 x \Phi(x) R + \cdots $. If the $d=2$ theory arises from a Kaluza Klein reduction of a higher dimensional theory, then $\Phi$ is the 
volume of the compact manifold.  } In this case, the entropy $\Delta S$ must be {\em equal to\/} the modular energy $\Delta K$ in the null limit. To show this, we recall that the von Neumann entropy is analytically determined by the R\'enyi entropies. The $n$-th R\'enyi entropy is given by the expectation value of twist operators inserted at the two boundaries of the spatial slab. The approach to the null limit can thus be organized as an operator product expansion. We argue that, in the limit,  the only operators that contribute to $\Delta S$  have twist $d-2$; and that for interacting theories in $d>2$, there is only one such operator. This implies that $\Delta S$ becomes linear in the density operator, and hence~\cite{Cas13}
\begin{equation}
\Delta K-\Delta S\to 0
\label{eq-skintro}
\end{equation}
in the null limit.

The unique twist 2 operator is the stress tensor. This implies a second key result:
\begin{equation}
\Delta S = \frac{2\pi}{\hbar} \int d^{d-2}y \int_0^1 dx^+\, g(x^+)\, T_{++}(x^+,y)~.
\label{eq-dsgenintro}
\end{equation}
Together with Eq.~(\ref{eq-skintro}), this extends the validity of Eq.~(\ref{eq-dkgen}) to the interacting case: the modular energy is given by a $g$-weighted integral of the stress tensor.

These arguments do not fully determine the form of the function $g(x)$. For interacting conformal field theories with a gravity dual~\cite{Mal97}, we are able to compute $g(x)$ explicitly from the area of extremal bulk surfaces~\cite{RyuTak06,HubRan07}.\footnote{Note that the bound we prove concerns light-sheets in the interacting theory when it is weakly coupled to gravity, {\em not}\/ light-sheets in the dual bulk geometry.} For $d>2$ we find that $g$ differs from the free field case, Eq.~(\ref{eq-gfree}).

However, our proof~\cite{BouCas14} of the Bousso bound did not depend on Eq.~(\ref{eq-gfree}). Rather, it is sufficient that $g$ satisfies a certain set of properties. We show that these properties hold in the interacting case. In particular, the key property
\begin{equation}
% \frac{dg}{dx^+}\leq \frac{g}{x^+}
\left| \frac{dg}{dx^+} \right| \leq  1 
\label{eq-gcrucial}
\end{equation}
can be established by considering highly localized excitations and exploiting strong subadditivity. This completes the extension of our proof to the interacting case.

\paragraph{Outline} This paper is organized as follows. Sections~\ref{ope} and \ref{sec-g} contain the new results sufficient to prove the Bousso bound in the interacting case (in the weakly gravitating limit).  In Sec.~\ref{ope} we consider the light-like operator product expansion of the defect operators that compute the R\'enyi entropies. We derive Eqs.~(\ref{eq-skintro}) and (\ref{eq-dsgenintro}), thus recovering a key step in the free-field proof: the local form of the modular energy, Eq.~(\ref{eq-dkgen}). We further constrain the modular energy in Sec.~\ref{sec-g}, where we establish Eq.~(\ref{eq-gcrucial}) for interacting fields. All remaining parts of the proof extend trivially to the interacting case.

In Secs.~\ref{holoq} and \ref{sec-whyzero}, we explore our intermediate results for the entropy and modular energy on null slabs, which are of interest in their own right. In Sec.~\ref{holoq}, we compute the $\Delta S$ explicitly for interacting theories with a bulk gravity dual. This determines $g(x^+)$ for these theories. For $d>2$, we find that $g(x^+)$ differs from the free field result. The approach to the null limit is studied in detail for an explicit example in Appendix~\ref{bh}. 

In Sec.~\ref{sec-whyzero}, we examine the vanishing of the relative entropy in the null limit, $\Delta S=\Delta K$. This arises because the operator algebra is infinite-dimensional for any spatial slab, whereas no operators can be localized on the null slab. Any fixed operator is eliminated in the limit and thus cannot be used to discriminate between states. Appendix~\ref{toyy} illustrates this behavior in a discrete toy model.

In Sec.~\ref{sec-conclusions}, we summarize our results and discuss a number of open questions.

\section{Entropies for Null Intervals in Interacting Theories} 
\label{ope}
In this section, we will explore the properties of the entropy of a quantum field theory on a spatial slab in the limit where the finite dimension of the slab becomes light-like (null). We consider free and interacting conformal field theories with $d\geq 2$ spatial dimensions. (We will comment on the non-conformal case at the end.) For interacting theories in $d>2$, we will find that the entropy is equal to the modular Hamiltonian, and that both can be expressed as a local integral over the stress tensor.

\begin{figure}[t]
\begin{center}
\includegraphics[scale=1.]{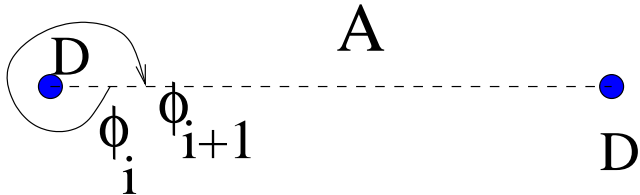}
\caption{The R\'enyi entropies for an interval $A$ involve the two point function of defect operators $D$  inserted at the endpoints of the interval. 
An operator in the $i^{th}$ CFT becomes an operator in the $(i+1)^{th}$ CFT when we go around the defect.}
\label{RegionDefect}
\end{center}
\end{figure}

It is convenient to consider the R\'enyi entropies first. The $n^\text{th}$ R\'enyi entropy $S_n(A) = (1-n)^{-1} \log \Tr \rho^n_A$ associated with a spatial region $A$ can be computed by taking the expectation value of a defect operator in a theory, which we denote by $\textrm{CFT}^n$, obtained from taking $n$ copies of a single CFT. The operator in question is a codimension 2 defect operator localized on the boundary $\partial A$ of a spatial region $A$ in the full Euclidean theory. In other words, the second orthogonal direction to the operator is Euclidean time. The defect operator is such that when we go around it, the various copies of the original CFT are cyclically permuted. In other words, an operator $\phi_k(x)$ defined on the $k^\text{th}$ CFT is mapped to $\phi_{k+1}(x)$ on the $(k+1)^\text{th}$ CFT, and $\phi_{n}(x)$ is mapped to $\phi_1(x)$; see Fig.~\ref{RegionDefect}.\footnote{These defect operators are oriented: there is a $D_+$ which maps $\phi_i \to \phi_{i+1}$ and a $D_-$ which maps $\phi_i \to \phi_{i-1}$.  For an interval, we have the insertion of $D_+$ and one end and of $D_-$ at the other end. We will not explicitly discuss this distinction.} This operator implements the boundary conditions for the replica trick \cite{CalWil94,CalCar09}.

To analyze the light-like limit, we start from the operators in Euclidean space. We then analytically continue them to Lorentzian time. Finally, we take the light-like limit. In this limit, we expect to have an operator product expansion. This expansion differs from the standard Euclidean operator product expansion in two respects. First, we are approaching the light-like separation, where the operators have zero metric distance but do not coincide, instead of approaching the coincident point along a purely space-like displacement. Second, in $d > 2$ dimensions, the two operators are extended and not local operators defined at a point. Despite these differences, we expect that there is a kind of operator product expansion that is applicable in this case. 

To our knowledge, the systematics of operator product expansions of extended operators in the light-like limit has not been explored. For the remainder of this section, we will make reasonable physical assumptions for the form of these operator product expansions. Operator product expansions for spacelike regions were considered in \cite{Hea10,Car13}.

First, we recall the form of the light-like operator product expansion 
for local operators. 
We will take the limit $x^2 \to 0$ with $x^+ \equiv x^0 + x^1$ held fixed. The expansion of two scalar operators has the form 
\begin{equation}
	 \label{opeexp} 
	 O(x) O(0) \sim \sum_k {|x|^{-2\tau_O + \tau_k}} (x^+)^{s_k} O_{k,s_k}.
\end{equation}
In this equation, the operator $O_{k,s_k}$ has spin $s_k$, scaling dimension $\Delta_k$ and twist $\tau_k \equiv \Delta_k - s_k$; and $\tau_O$ is the twist of the operator $O$. The twist governs the approach to the light-like limit. For finite $x^+$, we sum over all of the contributions with a given twist. 

In
free field theories, there are infinitely many higher spin operators with twist $d-2$. These operators contain two free fields, each with twist $\tfrac{1}{2}(d-2)$. In an interacting theory, all operators with spin greater than 2 are expected to have twist strictly larger than $d-2$. Furthermore, the twist is expected to increase as the spin increases \cite{Nac73} (see \cite{KomZoh12} for a more recent discussion).  The only operator with spin 2 and twist $d-2$ is the stress tensor, unless we have two decoupled theories. Operators with spin 1 include conserved currents. Scalar operators and operators with spin $1/2$ can have twist $\tau \geq \tfrac{1}{2}(d-2)$, with equality only for free fields.

As noted above, for $d>2$ the defect operators in question are extended along some of the spatial dimensions. We now discuss features of the operator product expansion in this case. Consider first the standard Euclidean OPE (as opposed to the light-like one). For such operators, the OPE is expected to exponentiate and become an expansion of the effective action for the resulting defect operator. In general, 
new light degrees of freedom could emerge when the two defect operators coincide. 
However, in our case 
the two twist operators 
annihilate each other, leaving only terms that can be written in terms of operators of the 
original 
theory. In other words, we expect 
\begin{equation}
	D(x) D(0) \sim \sum \exp 
		\left\{ 
			\int d^{d-2} y 
			\left[ 
				\sum_k \frac{1}{|x|^{d-2 -\Delta_k } } O_k( x=0, y)  
			\right] 
		\right\}
	\label{defope}
\end{equation}
where $y$ denotes the transverse dimensions and $O_k$ denotes local operators on the defect at $x = 0$. Thus the expansion is local in $y$. We can view this equation as an expansion of the effective action for the combined defect (consisting of both defects close together) by integrating out objects with a mass scale of order $1/|x|$.

The leading term in Eq.~\eqref{defope} is given by the identity operator and 
contributes
a factor of ${A_y}/{|x|^{d-2}}$ in the exponent (with a coefficient that depends on $n$), where $A_y$ is the transverse area. This is the expected form of $\Tr \rho_0^n = e^{-(n-1)S_n}$, which gives the vacuum R\'enyi entropies for the interval. In the vacuum case, all other operators have vanishing expectation values. 
This contribution cancels when we compute the difference $\Delta S$ of the von Neumann entropies of a general state and the vacuum, so we will not consider it further.

When we take the light-like limit of the R\'enyi defect operators, we expect to have an expansion which looks both like Eq.~\eqref{opeexp} and like Eq.~\eqref{defope}. In other words, we expect the expression to be local along the $y$ direction as in Eq.~\eqref{defope}, but with terms that are nonlocal along the $x^+$ direction 
as in Eq.~\eqref{opeexp}. 
In principle, along the $x^+$ direction, we can have terms which are very nonlocal. The operator $O_k(0,y)$ in Eq.~\eqref{defope} is replaced by an operator of the form on the right hand side of Eq.~\eqref{opeexp}:
 \begin{equation}
 \left. D(x) D(0) \right |_{\textrm{light-like}} \sim \exp
 \left\{ 
 	\int d^{d-2} y 
	\left[ 
		\sum_k   |x|^{  -(d-2)  + \tau_k  }   (x^{+})^{s_k} O_{ k ,s_k}   
	\right] 
 \right\}\,.
 \label{defopeli}
 \end{equation}

Note that the operators which appear in Eq.~\eqref{defopeli} are the operators of $\textrm{CFT}^n$ \cite{Hea10,Car13}. The generic form of these operators is \begin{equation}
 	O = O_1 O_2 \cdots O_n \,, 
	\label{opert}
 \end{equation}where $O_k$ is an operator on the $k^{\textrm{th}}$ copy of the original CFT. Some of the factors in Eq.~\eqref{opert} could be the identity, and the simplest operators we consider have only one factor which is not the identity. Performing the replica trick, the operators with a single factor that appear in the OPE of the two defect operators contribute to the entropy proportionally to an operator in the original CFT. Specifically, we find  \begin{equation}
	S_{\rm single} = \langle O_S \rangle\,. 
	\label{finaent}
\end{equation}Such contributions are linear in the density matrix, and therefore do not give rise to a non-zero value of $\Delta K - \Delta S$. The reason is that the operator on the right hand side is necessarily equal to $K$, since $K$ is the only operator localized to the region whose expectation value coincides with $\Delta S$ to linear order for any deviation from the vacuum state \cite{Cas13} (see also \cite{Won13}). 

\subsubsection*{The $d>2$ interacting case}

We will now argue that for interacting theories in $d>2$, all operators that contribute to Eq.~(\ref{defopeli}) are of this simple type: they all have only one nontrivial factor. In fact, only the stress tensor contributes.

Clearly, operators with $\tau > d-2$ will not contribute; this includes all higher spin operators in an interacting theory. Conserved spin 1 currents have twist $\tau = d-2$, but cannot appear because the defect operators are uncharged. Next, consider possible contributions from operators with twist $\tfrac{1}{2}(d-2) < \tau \leq d-2$. These operators could appear in representations which are not symmetric and traceless\footnote{Examples of such operators are fermion fields, or antisymmetric tensors in four dimensions. }. However, since the twist operator is invariant under transverse rotations, these operators must appear in pairs; their combined twist would be bigger than $d-2$.
  
Thus we can focus on the operators with spin zero. An operator of CFT$^n$ consisting of a single-copy scalar operator with twist in the range $\tfrac{1}{2}(d-2) < \tau \leq d-2$ would contribute to the entropy. This contribution will generically be divergent in the light-like limit to $\Delta S$, which is state dependent. In any case, single copy operators would give an equal contribution to $\Delta K$, so these operators do not contribute to $\Delta K - \Delta S$.\footnote{In some cases, these contributions are not present because of symmetry reasons. An example is the Wilson-Fisher fixed point at small $\epsilon = 4-d$. In this case, the dimension of $\phi$ is $\tfrac{1}{2}{ (d-2)} + O(\epsilon)$. However, due to the $\phi \to - \phi $ symmetry, this operator does not appear in the OPE of the defect operators involved in the replica trick.  Another example is the Klebanov-Witten theory \cite{KleWit98}. These are four dimensional theories with operators of dimension $3/2 < 2 $. However, these operators carry a $U(1)$ charge and cannot appear in this OPE. A relevant question here is whether there are theories with scalars with twists in this range which are not charged under any symmetry. If these operators are present, then our definition for $\Delta S$ will become divergent and will need to be modified.} On the other hand, if we had two operators in the range $\tfrac{1}{2}(d-2) < \tau \leq d-2$ on different CFT copies inside CFT$^n$, the total twist will be higher than $d-2$ and we will not get a contribution in the light-like limit.
     
This leaves the stress tensor, which has $\tau = d-2$ and can contribute in the null limit. However, unless $d=2$ (in which case $\tau=0$), only a single factor can contribute. Therefore, $\Delta S = \Delta K$ for interacting theories in $d>2$.

Notice that throughout this discussion, we have taken the coupling fixed and then taken the null limit. In particular, if we have a weakly coupled theory, we will get corrections to the result from free field theory which at each fixed order in perturbation theory will contain logs. One must resum the logarithms first, before taking the null limit, to recover the result that only the stress tensor survives.
      
Returning to the R\'enyi entropyies, we conclude that in interacting conformal theories, the only operator that can contribute to the expansion in the light-like limit is the stress tensor. 
All of its descendants contribute as well, so Eq.~(\ref{defopeli}) becomes a Taylor expansion around $x^+=0$. 
Discarding the contribution from the identity operator, which will drop out of $\Delta S$, we get
 \begin{equation}\label{config} 
 \left.D_n(x) D_n(0)\right|_{\textrm{light-like}}  \sim  \exp\left\{- (n-1) 2 \pi \int d^{d-2} y \int_0^{1} dx^+\, g_n(x^+) T_{++}(x^-=0,x^+, y) ] \right\}\,.
 \end{equation}
In this expression, we have set the size of the interval $\Delta x^+=1$ and extracted an overall factor of $n-1$ from the exponent. This factor accounts for the vanishing of the exponent for trivial R\'enyi operators when $n=1$. 
We have also replaced the sum over descendants by an integral over a  function, $g_n$,    determined by matching with a 
 Taylor expansion of the operator $T$. 
The integral is restricted inside the null interval because operators outside this range would not commute with the operators that are spatially separated from  
 the interval. 

The difference of von Neumann entropies of a general state and the vacuum is then given by analytic continuation:
 \begin{align}
 \Delta S &=\lim_{n\rightarrow 1} (1-n)^{-1}\log \langle D_n(x) D_n(0)\rangle \nonumber \\
 & =2 \pi \int d^{d-2} y \int_0^1 dx^+  g(x^+) T_{++}(x^-=0,x^+,y) \nonumber  \\
 &= \Delta K \, .\label{estrpc}
 \end{align}
The function $g$ is as yet undetermined and will be further discussed in the next section.

We expect the same holds for non-conformal theories with an interacting UV fixed point. For theories with a free UV fixed point, even if we expect that the modular Hamiltonian $K$ has the same general form in terms of the stress tensor, whether $\Delta K=\Delta S$ or not  would  generically 
depend on further details. For relevant deformations of a free UV fixed point we expect  to have $\Delta K\ge \Delta S$ as in the free theories, while we expect $\Delta K=\Delta S$ for asymptotically free theories.\footnote{In asymptotically free theories, the coupling runs as $g^2 \propto 1/\log \mu$ as a function of the scale $\mu$. The OPE is not given by a simple 
power behaviour but we need to integrate the anomalous dimensions of a range of scales as $ \exp[- \int { \frac{d\mu}{\mu} } \, \gamma(\mu)] $. Since 
$\gamma(\mu) \sim g^2(\mu) \propto 1/\log \mu  $, this integral diverges at short distances. Therefore, operators with non-zero anomalous dimensions do not contribute in the null limit, which involves going to very high scales. So we also expect Eq.~\eqref{estrpc} to hold.}

\subsubsection*{The case of free fields or $d=2$ interacting fields}

In free field theory, or if $d=2$, states with $\Delta S < \Delta K$ are known to exist on a null slab \cite{BouCas14}. We close this section by examining why the above argument for $\Delta S=\Delta K$ does not apply in these cases.
 
If the operator \eqref{opert} which appears in Eq.~\eqref{defopeli}  contains more than one nontrivial factor, it can give rise to a contribution to the entropy which is not equal to the expectation value of any operator in the original CFT. These contributions are interesting because they make $\Delta S < \Delta K$ possible. 
In a free field theory, such operators arise from insertions of the fundamental field $\phi$ in one copy and another field $\phi$ in another copy. They have twist $\tau=d-2$ and can contribute in the light-like limit.

In an interacting theory, all such operators gain a non-zero anomalous dimension. In particular, in a unitary theory, the field $\phi$ gains a positive anomalous dimension and so will not contribute in the null limit\footnote{In gauge theories, the fundamental fields are not gauge invariant on their own, and should be supplemented with Wilson lines as interactions are turned on. These Wilson lines end at the positions of the defect.}. 
However, in a $d=2$ interacting theory, multiple copies of the stress tensor can appear. Since $\tau = d-2=0$, the total twist will remain equal to $d-2$ no matter how many times the stress tensor appears in \eqref{opert}. Thus, in $d=2$, we can have  $\Delta S <  \Delta K$ even for interacting theories.

\section{Properties of $g(x)$ and Proof of the Bound}
\label{sec-g} 

In this section, we complete the proof of the Bousso bound by establishing sufficient properties of the function $g(x)$ in Eq.~\eqref{estrpc}. We begin with a list of simple properties that are expected on physical grounds. Near each boundary of the slab, the entanglement structure is like the origin of Rindler space, and $g(x^+) =x^+$ is known to hold for Rindler space from the Bisognano-Wichmann theorem \cite{BisWic76}.  Hence $g$ must satisfy
\begin{eqnarray}
g(0) & = & 0~, \nonumber\\
g'(0) & = & 1~, \label{eq-rindler}\\
g(1-x^+) & = & g(x^+)~,\nonumber
\end{eqnarray}
where the last relation arises from CPT symmetry. In the remainder of this section, we will derive the remaining property crucial to the proof of the bound: $|g'|\leq 1$.

Additional conditions arise from the theory of modular Hamiltonians. Let us define an operator on the global Hilbert space which we call the full modular Hamiltonian:
 \be
 \hat{K}_V=K_V-K_{V^c}\,,
 \ee 
where $V^c$ is the region complementary to $V$. For example, if $V$ is a Rindler wedge, then $\hat{K}_V$ is proportional to the boost generator. It is known that these Hermitian operators are monotonous under inclusion \cite{Bor95}, that is
 \be
 \hat{K}_V-\hat{K}_W\ge 0 \label{estp}
 \ee
 is a positive definite operator for any subregion $W\subseteq V$. This property can be seen as a consequence of monotonicity of relative entropy and strong subadditivity of the entropy \cite{BlaCas13}. 
 Let us first recall the definition of relative entropy $S(\rho|\rho_0) = \Tr[ \rho \log { \rho \over \rho_0} ] $. This is positive for any two density matrices. 
 Relative entropy can also be rewritten as 
 \begin{equation} \label{relentr} 
 S(\rho |\rho_0) = \Delta K - \Delta S
 \end{equation} 
  where $\Delta S = - \Tr[\rho \log \rho] + \Tr[\rho_0 \log \rho_0]$ and
 $\Delta K = \Tr[ \rho  K] + \Tr[ \rho_0 K  ] $ , with $K \equiv - \log \rho_0 + $constant. 
 The positivity of relative entropy implies that $\Delta S \leq \Delta K $. Now, the monotonicity of relative entropy is the following statement. 
Suppose we have two regions $W \subseteq V$ and we have two density matricies for the big region, $\rho_V$ and $\rho^0_V$. We can consider the restrictions
of these density matrices to the subregion $W$, call them $\rho_W$ and $\rho^0_W$. Monotonicity is the statement that 
$S(\rho_W |\rho^0_W) \leq S(\rho_V |\rho^0_V )$. 
 
   In the present case, we obtain two inequalities, one from $W\subseteq V$ and one from $V^c \subseteq W^c $
 \bea
 \Delta K_V-\Delta S_V&\ge& \Delta K_W-\Delta S_W\,,\\
 \Delta K_{W^c}-\Delta S_{W^c}&\ge& \Delta K_{V^c}-\Delta S_{V^c}\,.
 \eea
 where we have rewritten the relative entropies using (\ref{relentr}). 
Now we add these inequalities and separate the terms corresponding to the vacuum $\rho^0$ and the ones corresponding to a state $\rho^1$ 
 different from the vacuum. The terms involving entropy are \begin{equation}
 	S_V^0-S_{V^c}^0+S_{W^c}^0-S_{W}^0 = 0 \label{eq:1}
 \end{equation}
  which vanishes because the vacuum state is pure, and \begin{equation}
  	S_V^1-S_{V^c}^1+S_{W^c}^1-S_{W}^1 \ge 0 \label{eq:2}
  \end{equation}
  which is positive due to strong subadditivity\footnote{ The strong subaditivity statement we are using is $S(A) + S(B) \leq S(A \cup C) + S(B \cup C)$ where 
  $A,~B$ and $C$ are three disjoint systems. This property is sometimes also called weak monotonicity.}. The terms with modular Hamiltonians can be grouped into $\langle \hat{K}_V-\hat{K}_W\rangle^1$ and  $\langle \hat{K}_V-\hat{K}_W\rangle^0$. This last term is zero since the full modular Hamiltonian is a symmetry generator which annihilates the vacuum.\footnote{This property follows from the definition $K_V=-\log(\rho^0_V)$ and the Schmidt decomposition of the vacuum state across ${\cal H}_V\otimes {\cal H}_{V^c}$.} Hence we end up with the inequality
\be
\langle \hat{K}_V-\hat{K}_W\rangle^1\ge S_V^1-S_{V^c}^1+S_{W^c}^1-S_{W}^1\ge 0.
\ee  
This holds for any global state $\rho^1$ and implies (\ref{estp}).  

Going further, Eq.~\eqref{estp} implies the operator inequality
$K_V- K_W\ge K_{V^c}-K_{W^c}$ and hence
\be
\langle K_V\rangle ^1- \langle K_W\rangle ^1\ge 
\langle K_{V^c}\rangle^1-\langle K_{W^c}\rangle^1~.
\ee
Moreover, $\hat K_V |0\rangle =0$ implies $K_V^0=K_{V^c}^0$, and
similarly, $K_W^0=K_{W^c}^0$. Subtracting both of those equations, we now have
\be
\Delta K_V - \Delta K_W\ge \Delta K_{V^c} - \Delta K_{W^c}~.
\label{eq-kvkw}
\ee
In the null limit, this property is inherited by the full
modular Hamiltonians of null slabs. 

Now, let us consider a state whose stress-energy is positive and highly concentrated near some $x^+=\bar{x}^+\in W$. Such states can be produced by taking a fixed state and boosting it. We expect that in this limit the state outside the slab (in the region $W^c$, and hence also in $V^c$) is indistinguishable from the vacuum, so that $\Delta K_{V^c}\to 0$, $\Delta K_{W^c} \to 0$. For such states, Eq.~\eqref{eq-kvkw} reduces to
$\Delta K_V - \Delta K_W\ge 0$, and since both modular energies are positive, 
\be
	\frac{\Delta K_V}{\Delta K_W} \geq 1
\label{eq-kvkw2}
\ee

Now let $V$ be a slab with $x^+\in [0,1+\epsilon]$ and let $W \subset V$ be a slab with $x^+ \in [0,1]$. The modular Hamiltonian for slabs with non-unit affine length $\Delta x^+$ can be obtained by a simple coordinate transformation:
\begin{equation}
\Delta K = \frac{2\pi}{\hbar} \int d^{d-2}y \int_0^{\Delta x^+} d\hat{x}^+\, \Delta x^+\, g(\hat{x}^+/\Delta x^+)\, T_{++}(\hat{x}^+,y)~.
\end{equation}
Hence, the modular energies of the highly localized state satisfy \begin{equation}
	\frac{\Delta K_V}{\Delta K_W} = \frac{(1+\epsilon)\, g(\bar x^+/(1+\epsilon))}{g(\bar x^+)}\,.
\end{equation} In the limit as $\epsilon\to 0$, Eq.~(\ref{eq-kvkw2}) now implies
 \begin{equation}\label{monot} 
  \frac{dg}{dx^+}\leq \frac{g}{x^+}\, ,
  \end{equation}
Now, we repeat  the argument with  the region  $V$ as the rindler region with $x^+ \in [0, +\infty]$, and      $W$   the slab with $x^+ \in [0,1]$. 
For this region $V$ the function $g  =x^+$.  For a state with a concentrated stress tensor we obtain 
\begin{equation}
 { g(  x^+) \over x^+ }\leq 1
 \end{equation}
 Finally we conclude that
\be \label{boundder}
-1\le \frac{dg}{dx^+} \le 1\,. 
\ee 
where the first inequatity is obtained from the $g(x) = g(1-x)$ property (\ref{eq-rindler}). 

%and using Eq.~(\ref{eq-rindler}) we may conclude that
%\be \label{boundder}
%-1\le \frac{dg}{dx^+} \le 1\,. 
%\ee 

To prove the Bousso bound, we consider without loss of generality the null slab $x^+\in (0,1)$. We define $F(x^+) \equiv x^+ + g(x^+)$, which obeys $F(0)=0$ and $F(1)=1$ by Eq.~(\ref{eq-rindler}). We also have $F' \geq 0$ everywhere, by Eq.~(\ref{boundder}). These properties of the modular Hamiltonian suffice to show that the area difference along the light-sheet bounds the modular energy: ${ \Delta A \over 4 G_N} \geq \Delta K $ (see the discussion after eqn.\ 4.10 in Ref.~\cite{BouCas14}). As usual, positivity of the relative entropy implies that $\Delta S\leq \Delta K$ (with equality holding for $d>2$ interacting theories).  This completes the proof of the Bousso bound for both free and interacting theories, in the weakly gravitating limit.

\section{Holographic Computation of $\Delta S$ for Light-Sheets} 
\label{holoq} 
 
In this section, we consider interacting quantum field theories that have a gravity dual. In this case, the Ryu-Takayanagi formula  \cite{RyuTak06,HubRan07}  allows us to compute the entropy $\Delta S$ in the null limit. This will confirm our earlier demonstration that $\Delta S=\Delta K$, and will determine $g(x^+)$ explicitly for such theories. First, we consider a CFT; later, we will comment on the non-conformal case. Appendix~\ref{bh} discusses the approach to the null limit in greater detail.

We write the boundary metric as $ds^2 = - dx^+ dx^- + d {\vec y}^{\,2} $.  Let us first consider a spatial strip, extended along the $y$ directions.  One end of the interval is at $x^+ = x^- =0$ and the other end is at $x^+ = -x^- = \Delta x^+ $, a fixed constant. The bulk metric can be written as
 \be 
 ds^2 = { -  dx^+ dx^- + dy^2 + dz^2 \over z^2 }\,.
 \ee 
The minimal surface solution was found in \cite{BerCor98,RyuTak06-2}. It is given by 
\bea
 x^{+ } = - x^- &=  &  { \Delta x^+ \over 2 }  { u^d 
 F({1 \over 2} , { d \over 2(d-1) } , { 3d-2 \over 2(d-1)} ; u^{2(d-1)} )  \over F({1 \over 2} , { d \over 2(d-1) } , { 3d-2 \over 2(d-1)} ; 1 )}\,,
   \nonumber
 \\
 { z  } &= & { \Delta x^+\over 2} { u\,   d \over F({1 \over 2} , { d \over 2(d-1) } ,  { 3d-2 \over 2(d-1)} ;1 )}= z_{\max}\,u       \label{solutionxz}\,,
 \\ 
 {  A}_{\rm vacuum}   &= & A_{y} \int {\sqrt{  -   dx^+ dx^- + dz^2 } \over z^{d-1} } 
   \label{VacCon}\,,
 \eea
 where $F$ is the usual hypergeometric function,\footnote{Its values at $1$ can be written in terms of gamma functions: $F({1 \over 2} , { d \over 2(d-1) } ,  { 3d-2 \over 2(d-1)} ;1 )=\sqrt{\pi} d \,\Gamma(\frac{d}{2(d-1)})/\Gamma(\frac{1}{2(d-1)})$.} and $u\in (0,1)$ is a parameter describing the first half of the minimal surface, which is symmetric around $x^+=\Delta x^+/2$. The maximum $z_{\max}$ of $z$ is achieved for $u=1$. Here $A_y$ is the area in the $y$ directions. The formal expression for the area is UV divergent, but, as usual, we get a finite remaining contribution. 
 
We now consider a boosted interval. For that purpose we apply a combination of a boost in the $x^\pm$ plane and a dilation that transforms 
 \be \label{Scalings}
 x^+ \to x^+ ~,~~~~~~~x^- \to\eta^2 x^- ~,~~~~~~z \to \eta z  ~,~~~~~{\rm with} ~~~\eta \to 0\,.
 \ee 
This transformation takes the original spacelike interval to a null interval stretched along the $x^+$ direction. The proper length of the interval approaches zero. We also see that the surface is approaching the AdS boundary, in the sense that the largest value of $z$ is going to zero as $z \sim \eta \to 0$. Under these circumstances we find that the expression of the renormalized area (after subtracting the cutoff dependent piece) goes to minus infinity as $1/\eta^{ d-2}$. This is the expression for the vacuum entanglement entropy for the interval. 

Let us now consider a non-vacuum state. We expect that the minimal area surface will continue to approach the AdS boundary as we take the null limit. 
 Near the boundary, the metric approaches the AdS metric plus some small fluctuations. We can parametrize the metric as 
 \be
  ds^2 = { d z^2 + dx^{\alpha } dx^{\beta } ( \eta_{\alpha \beta} + h_{\alpha \beta } ) \over z^2 } ~,~~~~
  h_{\alpha \beta} \sim t_{\alpha \beta}(x) z^{d} + o(z^{d+1})\,.
  \ee
 Now the minimal surface action can be written as 
 \be 
 A = \int d^{d-2} y  { 1 \over z^{d-1} }  \sqrt{ -  dx^+ dx^- + dz^2 +  z^{d} \, t_{++}(d x^+)^2  }+ \cdots
 \ee
 where we wrote the part of the action that does not go to zero in the large boost limit, $\eta\to 0$. 
 More precisely, notice  that the first two terms inside the square root scale like $\eta^2$, while the last scales like $\eta^d$.
  We will assume that $d>2$ and return to the $d=2$ case later.  

\subsubsection*{The case of $d>2$}

For $d>2$, the last term in the 
 square root is a small perturbation and we can therefore expand the action.
  Due to the factor of  $1/z^{d-1} \sim 1/\eta^{d-1}$, the resulting first order term
 gives a finite answer 
 \bea
 A &=& A_{\rm vacuum } +  \int d^{d-2} y \,dx^+ \,{ z   \,  t_{++}  \over 
  2 \sqrt{ -   { dx^- \over dx^+ } + { d z^2 \over d{x^+}^2 }  } } \,,
 \\
 A -A_{\rm vac} &=&\frac{z_{\max}}{2} \int d^{d-2} y \,dx^+ \, t_{++}  u^{d}(x^+)  \,,\label{AreaDifference}
 \eea
 where the first term is the vacuum contribution in Eq.~\eqref{VacCon}. We have also used that the vacuum contribution is larger and determines the equations of 
 motion for the surface to the order we need in order to evaluate the second term. 
 We then see that the $g_{++}$ component of the metric gives a finite contribution. 
By performing a similar expansion, we can check that all other components of the metric do not contribute in the null limit either. For this, it is important to 
 use (\ref{Scalings}) to see how various terms behave. As an example, consider a component $h_{yy}$ in the metric. The component contains a $z^d \sim \eta^d $ which 
  multiplies the whole action that scales as $\eta^{  2-d}$. Since $d>2$, such a term does not contribute.   In a similar way, we discard higher orders in in the 
 expansion of the metric around $z=0$. 
 
\begin{figure}[t]
\begin{center}
\includegraphics[scale=1.]{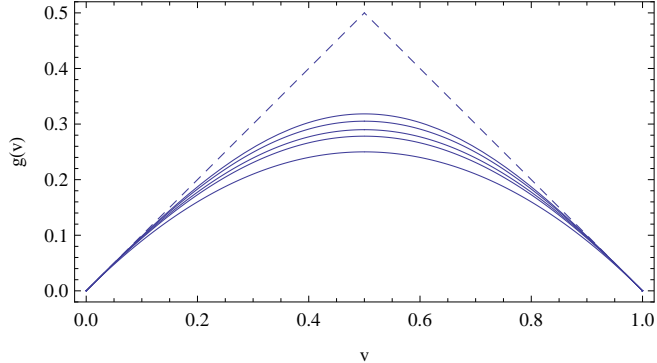}
\caption{The functions $g(v)$ in the expression for the modular Hamiltonian of the null slab, for conformal field theories with a bulk dual. Here $d=2,3,4,8,\infty$ from bottom to top. Near the boundaries ($v\to 0$, $v\to 1$), we find $g\to 0$, $g'\to \pm 1$, in agreement with the modular Hamiltonian of a Rindler wedge. We also note that the functions are concave (see Sec.\ 6). In particular, we see that $|g'|\leq 1$, in agreeement with our general argument of Sec.\ 3.}
\label{figu}
\end{center}
\end{figure}

 In conclusion,  the only part of the metric that matters is the first non-zero    term in the expansion of $h_{++}$. 
 This first-order term is also the term that gives the expectation value of the stress tensor, 
 \be
 t_{++} =\frac{16 \pi G_N}{d} \langle T_{++}\rangle \,,
 \ee 
 where $T_{++}$ is the value of the stress tensor (we have set the AdS radius to unity). 
 A similar expansion was performed in \cite{Cas13}.\footnote{ The authors of \cite{Cas13} considered a general surface and then expanded the metric to first order around the 
 AdS metric. Here, the argument is simpler because we only need the first order term in the 
 expansion of the metric near the boundary. Furthermore, we only need to consider the $g_{++}$ component.}  
 Using the solution   (\ref{solutionxz}), we can write (\ref{AreaDifference}) in the form 
 \be 
 \Delta S = { \Delta A \over 4 G_N} = 2 \pi\int d^{d-2} y \int_0^{\Delta x^+} dx^+\, \Delta x^+  g( x^+ / \Delta x^+)   \langle T_{++} (x^+,y,x^-=0) \rangle 
   \ee
   with $g$ defined parametrically by 
   \be \label{gresult}
   g( v) = {  u^d \over 2  F({1 \over 2} , { d \over 2(d-1) } , { 3d-2 \over 2(d-1)} ; 1 )}  ~, ~~~~~ v=  { u^d 
 F({1 \over 2} , { d \over 2(d-1) } ,  { 3d-2 \over 2(d-1)} ; u^{2(d-1)} )  \over 2 F({1 \over 2} , { d \over 2(d-1) } ,  { 3d-2 \over 2(d-1)} ; 1 )} \,. 
   \ee
   The function $g(v)$ is plotted for several dimensions in figure~\ref{figu}. Explicitly, we find $g(v)=v(1-v)$ for $d=2$, and in the limit $d\rightarrow \infty$ the function converges to $\sin(\pi v)/\pi$. For small $v$, we obtain the result $g(v)  = v + {\mathcal O}(v^2) $.

    We have thus obtained $\Delta S$ in terms of the expectation value of an operator, namely a certain integral of $T_{++}$. 
    According to the general argument discussed in Sec.~\ref{ope}, the operator in the right hand side is $\Delta K$; we obtain $\Delta S = \Delta K $. 

Notice that the relation $\Delta S = \Delta K$ gives values of the entropy on the light-sheet that are very different from naive expectations. For example, consider a thermal state. The entropy scales with the size of the interval as $(\Delta x^+)^2$, rather than the na\"ive (volume-extensive) entropy which grows like $\Delta x^+$ and which applies in the large temperature regime. Hence in this regime, we find $\Delta S$ is actually much greater than the naive entropy. To check in detail how the extensive entropy for spatial regions turns into a term that goes as $(\Delta x^+)^2$ for null surfaces, we have computed the areas of minimal surfaces in a black hole background. We find that there is actually a phase transition into a different class of extremal surface solutions as $\Delta x^-\rightarrow 0$. This is explained in more detail in Appendix \ref{bh}.   
    
    We can now briefly discuss the situation in non-conformal field theories. If we add a relevant deformation to the field theory, we are adding a scalar field 
    in the bulk which has a  profile  going like $ \phi \sim z^{\Delta }$ for small $z$. 
    This  affects the metric at quadratic order via terms of the form $\phi^2 \sim z^{ 2 \Delta }$. 
    Such terms modify only the diagonal components of the metric, and we have seen that as long as $ 2 \Delta > d-2$, such terms vanish. The latter is precisely the 
    unitarity condition for a non-free scalar operators. \footnote{ The unitarity condition is $ 2 \Delta \geq d-2$.  For equality, we have a
    free field in the boundary theory.}
   
\subsubsection*{The case of $d=2$} 
   
In two dimensions, it is still true that the minimal surfaces (geodesics) approach the boundary, but it is no longer true that we can treat the term involving 
$g_{++}$ in a perturbative fashion because it scales in the same way as the other terms. 
This implies that the final answer is non-linear in $T_{++}$. This non-linearity allows for $\Delta S < \Delta K$. 
   
For simplicity,   consider the special case of the theory at finite temperature (or in Rindler space).    Since it is related by a conformal transformation
to the plane, we can do all the computations explicitly by a simple coordinate transformation. 
The two point function of the twist operators is  
\begin{equation}
\langle \Phi_n(x) \Phi_n(0) \rangle =\frac{1}{  [ \sinh( \pi  { \Delta x^+ \over \beta } ) \sinh ( \pi  { \Delta x^- \over \beta }) ]^{ 2 \Delta_n } }
\end{equation}
with $x^\pm = \tau \pm \sigma $.
 This leads to the entropy \cite{CalCar09}
\begin{equation}
S = \frac{ c}{ 6 }
 \log \left( { \beta^2 \over \pi^2 \epsilon^2 }   \sinh ( \pi { \Delta x^+ \over \beta }) \sinh ( \pi  { \Delta x^- \over \beta } ) 
 \right)\,.
 \end{equation}
The vacuum case is given by the $\beta \to \infty$ limit, or $S = { c \over 6 } \log { \Delta x^+ \Delta x^- \over \epsilon^2 } $.
In the null  limit $\Delta x^- \to 0 $ we get
\begin{equation} \label{TwoDres}
\left.\Delta S \right|_{\Delta x^-=0}
= \frac{ c}{ 6} \log \left( { \beta \over \pi \Delta x^+ } \sinh ( \pi { \Delta x^+ \over \beta } ) \right)\,.
\end{equation}
This can be expanded as
\begin{align}
\Delta S &=  { c \over 6 } \left[ { x^2 \over 6 } - { x^4 \over 180 } + \cdots \right] ~,&~~~~~~~~~&x \ll 1 \,,\label{expans}\\
\Delta S &= { c\over 6 } \left[ x + {\rm constant} + \cdots \right] ~,&&x\gg 1\,,
\end{align}where $x = { \pi \Delta x^+ / \beta }$.
The first line is what we expect from the expansion of terms involving operators of the form
$ T_{++} , ~(T_{++})^2 $, possibly integrated at different points, replica copies,  etc. The last expression comes from resuming all
these operators. In this case, this agrees with what we expect from the operator product expansion, since all these operators
have twist zero in $d=2$. The important point is that operators on different replica copies survive the limit; see Sec.~\ref{ope}. 

Note that the modular Hamiltonian is 
\begin{equation}
\Delta K = 2 \pi\int_0^{\Delta x^+ } du\, u\, \left( 1- { u \over \Delta x^+} \right) T_{++} =  \frac{c x^2} {36}\,.
\end{equation}
since $T_{++} = \frac{c \pi}{12 \beta^{2}} $.
This agrees with the first term of the small $x$ expansion in (\ref{expans}), but in general it gives something larger than
$\Delta S$. This is particularly clear for $x \sim { \Delta x^+ \over \beta } \gg 1 $, and it can also be seen from the
quartic correction in (\ref{expans}). Therefore, in $d=2$, we get $\Delta S\le \Delta K$ but we do not get $\Delta S=\Delta K$.
   
Since all these results follow from conformal symmetry, it is clear that the gravity answer will reproduce them. This computation was done in \cite{RyuTak06-2}; one can check that the geodesics approach the boundary but Eq.~(\ref{TwoDres}) is reproduced.   
 
\section{Why is $\Delta S=\Delta K$ on Null Surfaces? }
\label{sec-whyzero} 

The relation $\Delta K=\Delta S$ is startling at first sight. It implies that the relative entropy between any state and the vacuum, $S(\rho_V|\rho^0_V)=\Delta K-\Delta S$, vanishes in the light-like limit. The relative entropy is a statistical measure of how easy is to differentiate between two states by making measurements. In general, the probability of confounding two states by making $N$ measurements falls off exponentially no faster than $e^{-N S(\rho_V|\rho^0_V)}$ (see e.g. \cite{Ved02}). If $\Delta K=\Delta S$, then the vacuum cannot be differentiated from any other state by making measurements of operators localized to a null surface. In other words, all states look the same as we approach this surface.

A related puzzle is the following: it is a general property of relative entropy that $S(\rho_V|\rho^0_V)=0$ implies $\rho_V=\rho^0_V$, but this in turn 
 would give $\Delta K=\Delta S=0$. However, the prediction $\Delta S = 0$ is not what we have found holographically.

\begin{figure}[ht!]
\begin{center}
\includegraphics[scale=1.2]{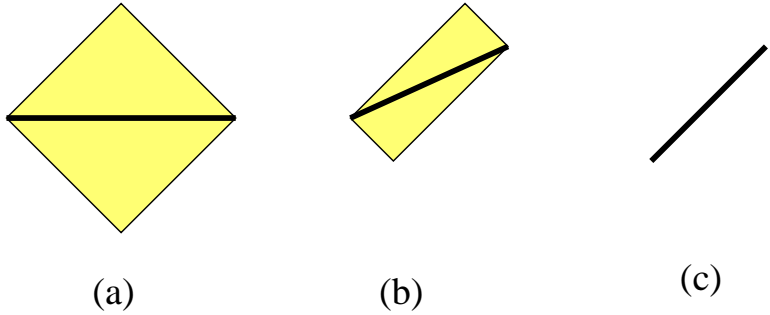}
\caption{Operator algebras associated to various regions. (a) Operator algebra associated to the domain of dependence (yellow) of a spacelike interval. (b) The domain of dependence of a boosted interval. (c) In the null limit, the domain of dependence degenerates to the interval itself.}
\label{boostedintervals}
\end{center}
\end{figure}
 
In this section, we explain both of these puzzles by noting that the quantities $\Delta S$ and $\Delta K$ are defined as limits for vanishing $\Delta x^-$. At finite $\Delta x^-$, the states are distinguishable by operators included in the algebra ${\cal A}(\Delta x^+,\Delta x^-)$ on the causal development of the spatial interval, but as we take the limit $\Delta x^-\rightarrow 0$, any fixed operator eventually drops out from the algebra. No operator remains in the intersection of all algebras, $\cap_{\Delta x^-} {\cal A}(\Delta x^+,\Delta x^-)$=1.\footnote{ Note that the operator $T_{++}$ evaluated on the null interval should not be considered as part of
the algebra because it sends states into non-normalizable states, even if we smear the operator along the null interval and the transverse directions. Nevertheless, the expectation value of this operator can be computed and can be different in two different states.} See figure \ref{boostedintervals}. 
 The same reason explains how $\Delta S$ and $\Delta K$ can be non-zero while the relative entropy is zero. This result cannot be correct for states on a fixed algebra, but it is a possibility for these quantities defined as limits on vanishing algebras. We describe how this can be accomplished using a toy model of an infinite chain of qubits in appendix \ref{toyy}.

In general, the relation $\Delta K = \Delta S \ne 0$ could not have been possible if the algebras for finite $\Delta x^{-}$ were not infinite dimensional. This phenomenon requires the full QFT, taking the UV cutoff to zero before 
 taking $\Delta x^-\rightarrow 0$. Otherwise, at finite $\Delta x^-$, we would run out of operators and find $\Delta K=\Delta S=0$.

Let us briefly describe what is meant by a full quantum field theory. A quantum field is an operator-valued distribution. In order to produce an operator acting on Hilbert space, a quantum field $\phi(x)$ has to be averaged by a smooth function of compact support $\phi_\alpha=\int d^dx \, \alpha(x)\phi(x)$. If $\alpha(x)$ is smooth on a $d$-dimensional spacetime region, we are guaranteed by the Wightman axioms that $\phi_\alpha$ is a well defined operator whose domain contains the vacuum state. The set of these operators where the support of $\alpha(x)$ is included in a spacetime region $V$ generates the algebra of operators acting in $V$.\footnote{These are von Neumann algebras. There is a technical point in that these algebras are better described as algebras of bounded operators. Bounded operators can be obtained from $\phi_\alpha$ by taking the projectors in its spectral decomposition \cite{Haa92}.}  

We want to see if a fixed localized operator can be defined for the null plane such that it remains in the intersection of the algebras $\cap_{\Delta x^-} {\cal A}(\Delta x^+,\Delta x^-)$ which implement the null limit. 
The problem of whether the domain of the test function $\alpha$ can be reduced to a spatial region or a region on a null plane, as opposed to a spacetime region, was studied in the past \cite{Ste63,Schl72}, mainly in attempts to develop a precise mathematical foundation to the usual canonical formalism of equal time commutation relations (see also \cite{Wal11}).  If $\phi_\alpha$ is a well-defined operator, we should have
\begin{equation}
\| \phi_\alpha|0\rangle\|^2=\langle 0|\phi_\alpha^\dagger \phi_\alpha|0\rangle=\int dx\, dy\, \alpha(x)^* \alpha(y)\langle 0|\phi(x)^\dagger \phi(y)|0\rangle< \infty\,.\label{23}
\end{equation}
This condition on the two point function of the field constrains its ultraviolet behavior. 

%Let us assume we have a CFT. 
%We are only interested in transformation properties of the fields under boosts in the $x^0,x^1$ coordinates, $(x^+,x^-)\rightarrow (\lambda \,x^+, \lambda^{-1}\,x^-)$. We classify the 
%operators according to spins and under behavior under these transformations. We take field components $\phi(x)$ that transform homogeneously as $\phi(x)\rightarrow \lambda^{s_+-s_-} \phi(x^\prime)$ under these boosts, with the actual spin of the field being $s\ge s_++s_-$. Here $s_+,s_-\ge 0$ are quantized as integers or half integers in the same way as $s$. 
%Suppose $\phi(x)$ is an operator of spin $s$ and dimension $\Delta$. The correlation function for spatial distances have a form fixed by the symmetries
%\begin{equation}
%\langle 0|\phi(x)^\dagger \phi(0)|0\rangle=c\, \frac{\,y^{2s-2s_-2s_-}(x^+)^{2 s_+}(x^-)^{2 s_-} }{(x^+x^-+y^2)^{\Delta+s}}\,,\label{fou}
%\end{equation} 
%where $y^{2s-2s_-2s_-}$ represents some homogeneous tensor on the transversal coordinates with order $2s-2s_+-2s_-$. 
%The Euclidean function has a similar form, which can be written in terms of derivatives of the distribution $(x^2)^{-\Delta+s}$. By analyticity this fixes the distribution %completely even at the light cone. In momentum representation this is proportional to
The Fourier transform of the two point function with no time ordering     $\langle \phi^\dagger \phi \rangle $ is 
\begin{equation}  
   \theta(p_0)  \theta(- p^2)   \frac{p_y^{2s-2s_+-2s_-} p_+^{2 s_+}p_-^{2 s_-}}{(p_+ p_- +p_y^2)^{\frac{d}{2}-\Delta+s}}\,, \label{fgls}
\end{equation}
where $p_y$ is a polynomial in the transverse components. 
To evaluate Eq.~\eqref{23}, we take the Fourier transform:
\begin{equation}
\alpha(x)=\int dp_+\,dp_- \,dp_y^{d-2}\, e^{-i(p_+ x^++p_- x^-+p_y y)} \alpha(p_+,p_-,p_y)\,.
\end{equation}  
For $\alpha(x)$ with support on the surface $x^-=0$, we have 
\begin{equation}
\alpha(p_+,p_-,p_y)= \alpha(p_+,p_y)\,,
\end{equation}
with $\alpha(p_+,p_y)$ independent of $p_-$ and falling off to zero faster than any polynomial in $p_+$ and $p_y$ due to the smoothness of $\alpha(x)$.
We have
\begin{equation}
\| \phi_\alpha|0\rangle\|^2\sim \int_{p^2<0,p_0>0} dp_+\,dp_- \,dp_y^{d-2}\,\frac{p_y^{2s-2s_+-2s_-}\,p_+^{2 s_+}p_-^{2 s_-}|\alpha(p_+,p_y)|^2}{(p_+ p_- +p_y^2)^{\frac{d}{2}-\Delta+s}}\label{conver}\,.
\end{equation}
The test function makes the integral convergent for large $p_+$ and $p_y$. However, the
   integral may not converge for large $p_-$. The best chance we have for it to converge is when $s_-=0$. 
   Power counting gives a convergent integral if 
\begin{equation}
\tau=\Delta-s<\frac{d-2}{2}\,, \label{inee}
\end{equation}
which is never the case for a unitary theory. 

For free fields, we have $\Delta-s=\frac{1}{2}(d-2)$, but the field obeys the wave equation so that instead of Eq.~\eqref{fgls}, we 
have an expression localized on the mass shell $p^2=-p_- p_++p_y^2=0$. In this case, the  denominator in Eq.~\eqref{fgls} is replaced by the delta function $\delta(p^2)$. Eliminating $p_+$ gives
\begin{equation}
\| \phi_\alpha|0\rangle\|^2\sim \int dp_- \,dp_y^{d-2}\,\Theta(p_-+p_y^2/p_-)\frac{p_y^{2s-2s_+-2s_-}\,(p_y^2/p_-)^{2 s_+}p_-^{2 s_-}|\alpha(p_y^2/p_-,p_y)|^2}{p_-}\label{convi}\,.
\end{equation}
This integral in $p_-$ is logarithmically divergent for a free scalar field with $s_+=s_-=0$, but converges for $\partial_+ \phi$ and its derivatives with $s_+>s_-$. For a free spin $1/2$ field, it converges for the $\psi^+$ component (and derivatives). For a Maxwell field tensor $F_{\mu\nu}$, we must again take the component with $s_-=0$ and $s_+=1$, that is, the components $F_{y,+}$.

So only for free fields do we expect to have operators localized on the null surface and $\Delta S\neq \Delta K$ for general states. The localized operators can be non-local in the $y$ direction so that $\Delta S$ does not need to decompose into a sum of contributions from each of the null lines. 

For non-conformal theories with a free UV fixed point, the localizability of the operators depends on the details of the approach to the fixed point. Using the spectral representation of the two point function for a scalar field in terms of that of a free massive scalar field 
\begin{equation}
\langle 0|\phi(x)^\dagger \phi(0)|0\rangle=\int dm^2\, \rho(m^2)\, G_0(x,m^2)\,,
\end{equation}
the general result \cite{Schl72} is that the derivative $\partial_+ \phi$ of this scalar field can be localized only if
\begin{equation}
\int dm^2\, \rho(m^2)<\infty\,.
\end{equation}   
This condition gives a finite wave function renormalization, which is expected to hold for superrenormalizable theories but not for marginal renormalizable theories \cite{Str75}.

\section{Conclusions} 
\label{sec-conclusions}

\paragraph{Summary.} 

We explored some properties of the entropy associated to null slabs in general interacting field theories.  We found a general expression for the modular Hamiltonian in terms of a local integral of the stress tensor components along the null slab, Eq.~\eqref{eq-dkgen}. We derived this by considering the light-cone OPE for the defect operators that compute the R\'enyi entropies; general arguments involving the spectrum of operators then constrain the von Neumann entropy and show that it is equal to the modular Hamiltonian.  

We also proved certain inequalities obeyed by the function $g$ that multiplies the stress tensor in the modular Hamiltonian. These inequalities, Eqs.~\eqref{monot} and \eqref{boundder}, were previously shown~\cite{BouCas14} to be sufficient for the Bousso bound~\cite{CEB1}. Our work extends our earlier proof of the Bousso bound to interacting theories.

We computed the entanglement entropy in the null limit for theories with a gravity dual. In the null limit, the minimal surface approaches the AdS boundary. The change in the area can be found from the asymptotic form of the metric. This asymptotic form of the metric also determines the stress tensor. Therefore, we get a result that is in line with our general expectations. We view this as an additional consistency test on the holographic entanglement entropy formula~\cite{RyuTak06,HubRan07} in a strongly Lorentzian context. Our analysis fully determines the function $g$ for such theories, Eq.~\eqref{gresult}, and it shows that $g$ takes a different form than in the free theory.

A curious feature of our result is that, for interacting theories, the change in entropy is exactly given by the change in the expectation value of the modular Hamiltonian: $\Delta S = \Delta K$. In a finite-dimensional Hilbert space, this relation would also imply that both $\Delta S$ and $\Delta K$ are zero. Here, however, they are non-zero.  This is possible because we are taking a limit that involves infinite dimensional algebras. We also saw that no elements remain in the algebra after we take the limit. One can still consider limiting values of expectation values of operators on the null line, but such operators, or their smeared versions on the null surface, do not define reasonable operators on the Hilbert space because their variance is infinite. Physically, this result means that in interacting theories, one cannot distinguish between any two states by making measurements purely on the light-sheet. Appendix B presents a simple toy model involving an infinite number of qubits where similar features are present.

\paragraph{Discussion and open problems.} 

The Bousso bound involves the notion of an entropy flux through the light-sheet. Defining a local notion of entropy current is notoriously difficult in quantum field theory. Here we have defined it through $\Delta S$, the difference in the von Neumann entropies of the interval between two different quantum states.  This notion does indeed have properties that suffice to ensure the validity of the corresponding Bousso bound.  Nevertheless, the quantity $\Delta S$ has some counter-intuitive properties. 

The most surprising aspect of this definition is that we find $\Delta S = \Delta K$, which means that all ordinary states are indistinguishable by local measurements on the light-sheet. We have not found more familiar-looking definitions for the entropy flux, to which a Bousso-type bound might apply. Further research will be needed to better understand the relation between $\Delta S$ and more conventional (spacelike) definitions for the entropy flux.

Notice that the energy flux is given by a local quantity, the expectation value of $T_{++}$. On the other hand, $\Delta S$ is non-local since the function $g$ depends on the positions of the endpoints of the interval. Thus, it cannot be viewed as the flux of a local operator.

We expect that $\Delta S $ will provide an upper bound to the more familiar concepts of entropy. For example, in a theory where we can define an entropy current, as in hydrodynamics, we expect that $\Delta S$ should be larger than the flux of the entropy current on the light-sheet.  In the holographic computations involving black branes, this is indeed true. The reason is very simple: the entropy flux scales like the length of the interval, $\Delta x^+$, on the other hand $\Delta S$ scales like $(\Delta x^+)^2$. The relative coefficient involves the temperature, $T$. This means that if $\Delta x^+ $ is somewhat greater than $\beta = 1/T$, then $\Delta S$ will be larger than the entropy flux.  We also see this clearly in the two dimensional results, Eq.~(\ref{TwoDres}). We expect this to be a general feature of thermal or hydrodynamic states.
% Thus, from the fact that $\Delta K$ scales quadratically in $\Delta x^+$ we expect that for general theories 
%$\Delta S$ will be larger than the hydrodynamic entropy flux as long as $\Delta x^+ $ is somewhat greater than the inverse temperature. 

An interesting conclusion is that information in interacting theories becomes very delocalized on the light front. Information that is fairly localized along the longitudinal direction in free theories spreads once we include interactions. We also expect that the mutual information between a null interval and any other fixed region should vanish. This follows from the result $\Delta S = \Delta K$.  We also see this in the holographic examples. 
% On the other hand, it is surprising if we want to interpret $\Delta S$ as an extension of the notion of an entropy flux. 

In a CFT with a gravity dual, the entropy $\Delta S$ for spatial slabs in a thermal state displays a phase transition as the null limit is approached (see Appendix~\ref{bh}). This is likely to hold in general for states which start out with a non-zero $\Delta S$ for a spacelike interval in the large $N$ approximation. 
 
For free theories, one can prove not only the covariant bound but the stronger result of monotonicity~\cite{BouCas14}: ${ \Delta A \over 4 G_N} - \Delta S$ never decreases under inclusion in a larger light-sheet. This follows from the concavity of the function $g$, $g''<0$, which holds in the free case. Here, we found that this property continues to hold for interacting theories with a holographic dual (Fig.~\ref{figu}), so monotonicity of ${\Delta A \over 4 G_N} - \Delta S$ follows in these cases. We leave a general proof of $g''<0$ to future work.
%$Regarding the function $g$ appearing in the modular Hamiltonian we can note the following. In a free theory, $g$  was the same as the one in $d=2$, 
%$g(x) = x(1-x)$, which could be viewed as a reduction of the higher dimensional field theory into two dimensional modes on the light-sheet. So, when we have
%a different $g$, we seem to have a failure of this notion and it would be interesting to understand futher how it happens. 
It would also be nice to compute the function $g$ to first order in perturbation theory for a weakly coupled CFT, such as ${\cal N}=4$ super Yang-Mills.

 \bigskip

\acknowledgments 

We thank D.\ Marolf and A.\ Wall for discussions. R.B.\ and Z.F.\ are supported in part by the Berkeley Center for Theoretical Physics, by the National Science Foundation (award numbers 1214644 and 1316783), by the Foundational Questions Institute grant FQXi-RFP3-1323, by ``New Frontiers in Astronomy and Cosmology'', and by the U.S.\ Department of Energy under Contract DE-AC02-05CH11231.  H.C.\ thanks the Institute for Advanced Study for hospitality and financial support. H.C.\ is partially supported by CONICET, CNEA, and Univ.\ Nac.\ Cuyo, Argentina. J.M.\ is supported in part by U.S.\ Department of Energy grant DE-SC0009988.

\appendix

\section{Extremal Surfaces and Phase Transitions on a Black Brane Background}
\label{bh}

In this appendix, we consider a thermal state in an interacting CFT with a bulk dual, an asymptotically anti-de Sitter planar black brane spacetime. This allows us calculate the entanglement entropy holographically using the HRT prescription  \cite{HubRan07}. We are able to study the approach to the null limit in detail, showing that the entropy on sufficiently large slabs undergoes a phase transition at large boost. We reproduce the result $\Delta S = \Delta K$ for the null slab. 

The metric of a black brane in AdS is
\begin{equation}\label{bhmetr}
ds^2 = \frac{  -f \, dt^2  + dx^2 + dz^2/f + dy^2}{z^2} ~,~~~~~~~~f = 1 - z^d/z_0^d\,.
\end{equation}
The inverse black hole temperature is
\begin{equation}
\beta=\frac{4 \pi z_0}{d}\,,
\end{equation}
and the energy density is given by
\begin{equation}
T_{00}=\frac{(d-1)}{16 \pi G_N}\frac{1}{z_0^d}\,.
\end{equation}
It follows that the null-null component of the stress tensor is
\be 
T_{++}=\frac{d}{4(d-1)}T_{00}=\frac{d}{64 \pi G_N}\frac{1}{z_0^d}\,.\label{fiyo}
\ee
The extremal surface action is
\begin{equation}\label{lagra}
I = \int d z \, \frac{ 1 }{ z^{d-1} } \sqrt{ - f \, \dot t^2 + \dot x^2 + 1/f}\,.
 \end{equation}
 Let the momentum conjugate to $x$ be denoted $p$. We find
 \begin{equation}\label{solutc}
 p = \frac{1}{ z^{d-1} } \frac{ \dot x }{  \sqrt{ - f\,\dot t^2 + \dot x^2 + 1/f} } ~,~~~~~~~~~~~
 \dot x = p  z^{ d-1} \frac{ \sqrt{ 1/f - f\,\dot t^2 }}{ \sqrt{ 1 - p^2 z^{ 2 (d-1)} } }\,.
 \end{equation}
 Define a new effective Lagrangian $L' \equiv L - p \dot x $:
 \begin{equation}\label{newla}
 L' =  \frac{ 1 }{ z^{d-1} } \sqrt{ 1/f - f \, \dot t^2 } \sqrt{ 1 - p^2 z^{ 2 ( d-1)} } = \sqrt{ { 1 \over z^{ 2 (d-1)}} - p^2   }
  \sqrt{ 1/f -f\, \dot t^2 }\,.
  \end{equation}
Writing $E$ for the momentum conjugate to $t$, we obtain
\begin{eqnarray}\label{obtaim} 
E  &=&   \frac{ f\, \dot t  }{  \sqrt{ 1/f - f\,\dot t^2}  \sqrt{ { 1 \over z^{ 2 (d-1)}} - p^2   }} \,,
\\
\dot t &=&    \frac{ E  }{ f^{3/2}  \sqrt{ E^2/ f  + { 1 \over z^{ 2 (d-1)}} - p^2  }}\,\label{obtaimT} ,
\\
\dot x &=& \frac{ p   }{ \sqrt{f}  \sqrt{ E^2/ f  + { 1 \over z^{ 2 (d-1)}} - p^2  }}\,\label{obtaimX} .
\end{eqnarray}
These are the equations of motion of the extremal surfaces. We take $E,p>0$ and fix scale invariance by setting $z_0=1$ in the function $f(z)$, so $f(z) = 1 -z^d$. Integrating these trajectories, we obtain the null coordinates $\Delta x^\pm$
of the the extremal surface solutions at the boundary,
\begin{equation}
\Delta x^\pm=2\int_0^{z_r} dz\, \left({  E \over f }\pm p  \right) \frac{ 1 }{ \sqrt{f} \sqrt{ E^2/f + { 1 \over z^{ 2 (d-1)}} - p^2  } }\,.\label{deltas}
\end{equation}
We can also rewrite the initial action (i.e. area) from Eq.~\eqref{lagra} as 
\begin{equation}
\label{sola} 
I = 2\int_0^{z_r} dz\, \frac{ 1 }{ z^{ 2(d-1)} \sqrt{f} \sqrt{ E^2/f + \frac{ 1 }{ z^{ 2(d-1)}} - p^2 } }\,.
\end{equation}
In these integrals, the upper limit $z_r$ is the return point of the trajectory, which is the smallest positive root of the denominators of Eqs.~\eqref{obtaimT} and \eqref{obtaimX}; that is,
\begin{equation}
 \frac{E^2}{1-z_r^d}  + \frac{ 1 }{ z_r^{ 2 (d-1)}} - p^2=0\,.\label{max}
\end{equation}

The turning point is calculated from the smallest solution of this equation, with $z_r\in (0,1)$.  
The first two terms in Eq.~\eqref{max} are positive when $z\in (0,1)$, and the second is greater than one. We conclude that an extremal surface which returns to the boundary exists for all $p>1$. As we increase $E$ from zero with $p > 1$, there are solutions for $z_r$ only up to a maximum value of $E$, which we denote by $E_{\textrm{max}}(p)$. This maximum value of $E$ is simultaneously the solution to Eq.~\eqref{max} and 
\begin{equation}
\frac{d}{d z_r}\left(\frac{E^2}{1-z_r^d}  + { 1 \over z_r^{ 2 (d-1)}}\right)=0\,.
\end{equation}
When $E>E_{\textrm{max}}$, there are no extremal surface solutions that return to the AdS boundary. 

\begin{figure}[t]
\begin{center}
\includegraphics[scale=1.]{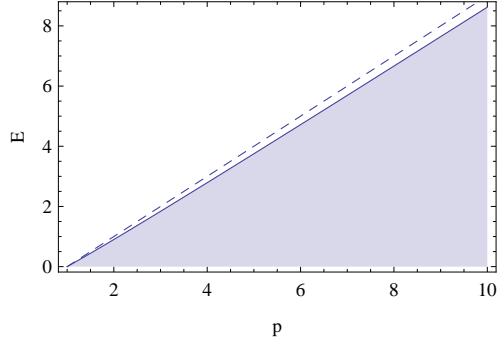}
\caption{The maximum value $E_{\textrm{max}}(p)$ of $E$ for getting a surface that returns to the boundary (solid line). For comparison, we also plotted the line $E=p-1$ (dashed line). The extremal surface solutions of interest appear in the region $p>1$, $0<E<E_{\textrm{max}}(p)$. Here, we have taken $d=3$.}
\label{emax}
\end{center}
\end{figure}

In Fig.~\ref{emax}, we plot the parameter space $(p,E)$ for $d=3$. In $d=3$, the curve $E=E_{\textrm{max}}(p)$ runs near the line $E=p-1$; the plot looks similar in other dimensions. The shaded region contains the extremal surface solutions.

Interestingly, the extremal surface for $E=E_{\textrm{max}}(p)$ corresponds to a time-like region on the boundary, with $\Delta t>\Delta x$. This counter-intuitive result is possible because even if the boundary interval is time-like, we are still considering locally spatial surfaces in the bulk. However, these extremal surfaces cannot be regulated by vacuum-subtraction, because the extremal surface solutions with this boundary region in vacuum AdS do not have a well-defined area. 
So the parameter space we are interested in is further reduced to $E<E_{\textrm{null}}(p)\le E_{\textrm{max}}(p)$, where $E_{\textrm{null}}(p)$ denotes the energy for which the extremal surface solution has $\Delta x^-=0$. 

\begin{figure}[t]
\begin{center}
\includegraphics[scale=.7]{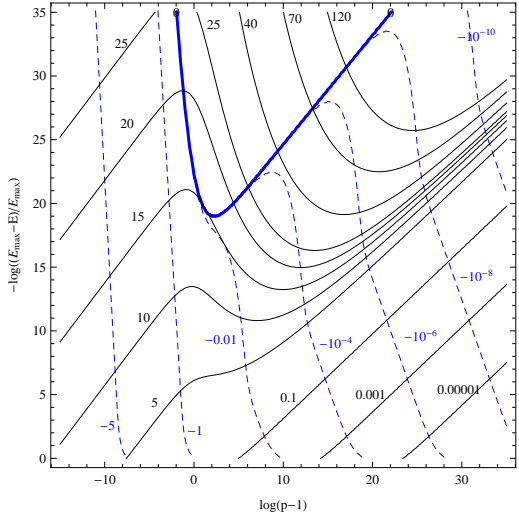}
\caption{Curves of constant $\Delta x^+$ (black solid curves) and $\Delta x^-$ (blue dashed curves), in the logarithmic parameter space defined by $(\log(p-1),-\log(E_{\textrm{max}}(p)-E)/E_{\textrm{max}}(p))$. The value $p = 1$ maps to $-\infty$ and $p = \infty$ maps to $+\infty$ on the horizontal axis, while $E = 0$ maps to 0 and $E = E_\textrm{max}(p)$ maps to $+ \infty$ on the vertical axis. The thick blue contour represent the null solutions with $\Delta x^-=0$. Above this contour, the boundary interval is time-like.  If $\Delta x^+\gtrsim 15$ and we follow a contour of constant $\Delta x^+$, we find two solutions with exact $\Delta x^-=0$. For all contours of fixed $\Delta x^+$, there exists an asymptotic null solution in the limit $p\rightarrow \infty$.}
\label{contour}
\end{center}
\end{figure}

The separation between $E_{\textrm{max}}(p)$ and $E_{\textrm{null}}(p)$ in parameter space is, however, exponentially small. We show the relevant contour in figure in Fig.~\ref{contour} in logarithmic variables (for $d=3$; other dimensions are similar).  

Numerical analysis of the solutions shows that, for $\Delta x^+\sim 1$ and smaller\footnote{Recall we have set $z_0$ to unity.}, there are no exact solutions with $\Delta x^-=0$, only an asymptotic set of solutions for which $\Delta x^+$ is fixed and $\Delta x^-$ approaches but never exactly reaches zero. The parameters $p$ and $E$ go to infinity in the limit $\Delta x^- \rightarrow 0$, and the extremal surface runs closer to the AdS boundary. We call this family of solution the ``perturbative solutions,'' because $\Delta S$ can be computed perturbatively in this case (see Sec.~\ref{holoq}). For sufficiently large $\Delta x^+$ (larger than approximately $15$ in $d=3$), in addition to the asymptotic solution, there exist two other solutions with finite $E_{\textrm{null}}(p)$ such that $\Delta x^-=0$ exactly. Fig.~\ref{contour} gives the contour plot of $\Delta x^+$ and $\Delta x^-$ for a region of the parameter space $(p,E)$. We plot the solutions in logarithmic parameter space in Fig.~\ref{contour}. Following a contour of constant and sufficiently large $\Delta x^+$ from left to right in this diagram, $\Delta x^+\gtrsim 15$, we intersect the contour $\Delta x^-=0$ twice, corresponding to the two precisely null solutions. The part of the contour to the left of the first intersection (with $p \sim 1$) is the ``thermal'' family of solutions. These solutions have a thermal character because most of the surface extends near to the horizon of the black hole. Hence the entropy contains a term that grows like $\Delta x^+$, a volume-extensive term that goes with the thermal entropy density. By increasing $p$ along a contour of fixed $\Delta x^+$, we again approach the asymptotic perturbative solution.

Let us see in more detail how the area behaves in these two solutions. As shown below, there is a third null solution, but it has greater area than the other two and, according to the HRT prescription, should not be regarded as the entropy.

\subsection*{Perturbative solution}

According to Sec.~\ref{holoq}, we expect the limiting value of the entropy to take the form 
\begin{equation}
\Delta S=2\pi\,A_\perp\,  (\Delta x^+)^2 T_{++} \int_0^1 dv\, g(v)\,. 
\end{equation}
The difference between the perturbative extremal surface area and the vacuum area is 
\begin{equation}
\Delta A=8 \pi G_N \,A_\perp\,  (\Delta x^+)^2 T_{++} \int_0^1 dv\, g(v)\,.
\end{equation}
 Using Eq.~\eqref{fiyo} and the explicit form of the function $g(v)$, we obtain
\begin{equation}
\Delta A=\frac{\Gamma\left(\frac{d}{d-1}\right)\Gamma\!\left(\frac{1}{2(d-1)}\right)^2}{32 \pi^{1/2}(d-1)\,\Gamma\!\left(\frac{3d-1}{2(d-1)}\right)\Gamma\!\left(\frac{d}{2(d-1)}\right)^2}\frac{A_\perp\,(\Delta x^+)^2}{z_0^d} \,.\label{perturbative}
\end{equation}
Setting $A_\perp=z_0=1$, we obtain perfect accord with our numerical simulation of the extremal surfaces.

\subsection*{Thermal solution}

This solution captures the thermal entropy. We expect the difference in extremal surface areas to approach $\Delta x^+/2$ asymptotically at large $\Delta x^+$. 

The thermal solutions track the horizon of the black hole at $z_0 = 1$. In parameter space, this occurs when $E$ is of the same order as $(p-1)$. When this is the case, most of the contribution to the integral comes from the region where $z$ is order $z_0 = 1$, and we can expand the integrand around that point. 

First we perform the substitutions
\begin{equation}
 p = 1+ 2 \delta \epsilon ~,~~~~~~~~~z = 1 - u \epsilon/(d-1)\,,   ~~~~~~~~~~ E = \sqrt{ \delta^2 -\sigma^2} \sqrt{\frac{ 2 d }{ d-1}} \epsilon\,.
 \end{equation}
In this limit, the integrals become
\begin{eqnarray}\label{integr} 
\Delta x &=& { \sqrt{ d-1 \over 2 d }} \int_{\delta + \sigma } { du \over \sqrt{ ( u -\delta)^2 - \sigma^2 } }\,,
\\
\Delta t &=&  { (d-1) \sqrt{\delta^2 -\sigma^2 } \over d } \int_{\delta + \sigma}  { du  \over u \sqrt{ ( u-\delta)^2 - \sigma^2 }}\,,
\\
A_{ren} &=&  \sqrt{ d-1 \over 2 d } \int_{\delta + \sigma} { du \over \sqrt{ ( u -\delta)^2 - \sigma^2 } } = \Delta x =\Delta x^+/2\,.
\end{eqnarray}
The renormalized area is obtained by subtracting the divergent piece with a UV cutoff $\epsilon$.  
Note that the zeros in the denominator occur at $u = \delta \pm \sigma $, both of which we take to be positive.
Additionally, there is a zero at $u =0$ in the denominator of the integral for $\dot t$. The integral over $u$ starts at the largest zero, $u^*=\delta + \sigma$, and
moves to larger values of $u$ (which corresponds to smaller values of $z$). 
We can do these integrals and focus on the potentially large terms at small $\delta,\sigma$. We obtain
\begin{eqnarray}
\Delta x &=&  - \sqrt{ d-1\over 2 d } \log \sigma\,,
\\
\Delta t &=& - { d-1 \over d } \log\left[ \delta - \sqrt{ \delta^2  - \sigma^2 } \over \sigma \right]\,,
\\
A_{ren} &=& \Delta x\,.
 \end{eqnarray}
 The last equation implies that the entropy flux is what we expected. 
We take the ansatz $\sigma \sim \gamma \delta^a$ with $a>1$, where $\gamma$ is some constant. 
The expansions become, for small $\delta$,
\begin{eqnarray}\label{obtaint}
\Delta x &=&  - \sqrt{ d-1\over 2 d } a \log \delta\,,
\\
\Delta t &=& - { d-1 \over d } (a-1) \log \delta\,.
\end{eqnarray}
Setting $\Delta x = \Delta t$ for the null solution, we find
\begin{equation}\label{solva}
a = { 1 \over 1 - \sqrt{   d \over 2 (d-1)} }\,.
\end{equation}
This means that thermal solutions with exact $\Delta x^-=0$ exist for large $\Delta x^+$. 

However, we are interested not in the renormalized area but in the area difference with respect to the vacuum solution. Using the area for the vacuum solution \cite{RyuTak06-2}, the area difference for large $\Delta x^+$ is
\be
\Delta A\simeq \frac{\Delta x^+}{2} +\frac{2^{d-1}\pi^{(d-1)/2}}{d-2} \left(\frac{\Gamma(\frac{d}{2(d-1)})}{\Gamma(\frac{1}{2(d-1)})}\right)^{d-1} \frac{1}{(-\Delta x^+ \Delta x^-)^{\frac{d-2}{2}}} \,.\label{thermal}
\ee

\subsection*{Phase transition for large $\Delta x^+$}

Comparing Eqs.~\eqref{perturbative} and \eqref{thermal}, wee see that the perturbative solution has less area than the thermal one for sufficiently small $\Delta x^-$. This occurs because the perturbative solution has the same negative and finite term as the vacuum solution which grows like $(\Delta x^-)^{-\frac{d-2}{2}}$. Hence this term does not appear in the area difference in Eq.~\eqref{perturbative}. The thermal solution cannot have this term because it is an exact solution valid for $\Delta x^-=0$; its area cannot depend on $\Delta x^-$. Therefore, the area difference,  Eq.~\eqref{thermal}, diverges as $(\Delta x^-)\rightarrow 0$ for the thermal class of solutions. However, for finite values of $\Delta x^-$ and sufficiently large values of $\Delta x^+$, the thermal solution must have smaller area, since it increases only linearly with $\Delta x^+$ while the perturbative solutions grow quadratically. The phase transition occurs when the area of the two solutions becomes equal, which is approximately given by
\be
(\Delta x^+)^{\frac{d+2}{2}}(-\Delta x^-)^{\frac{d-2}{2}}=\frac{2^{d+4}\pi^{d/2}(d-1)}{d-2}\frac{ \Gamma\left(\frac{3d-1}{2(d-1)}\right)}{\Gamma\left(\frac{d}{d-1}\right)} \left[\frac{\Gamma(\frac{d}{2(d-1)})}{\Gamma(\frac{1}{2(d-1)})}\right]^{d+1}\,.
\ee

\begin{figure}[t]
	\centering
		\includegraphics[height=2.7in]{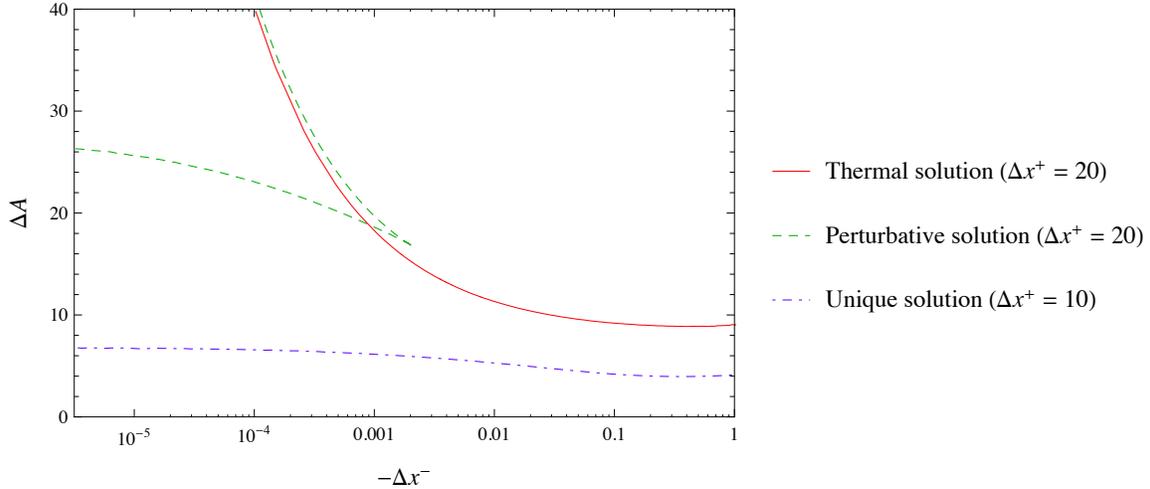}
	\caption{The vacuum-subtracted extremal surface area versus $\Delta x^-$ for fixed $\Delta x^+$ ($\Delta x^+=20$ and $\Delta x^+ = 10$ for $d = 3$ is shown). This numerical simulation demonstrates that, for sufficiently large $\Delta x^+$ (in $d=3$, the condition is $\Delta x^+ \gtrsim 15$), there exists a phase transition at finite $\Delta x^-$ to a different, perturbative class of solutions. At smaller $\Delta x^+$, there is no such phase transition.}
	\label{fig:dxp2010d3}
\end{figure}

We have numerically evaluated $\Delta A$ as a function of $\Delta x^-$ for fixed $\Delta x^+$ in $d=3$ dimensions. The result is shown in Fig.~\ref{fig:dxp2010d3}. We observe that, as predicted, the thermal solution tends toward infinite area as we take the limit $\Delta x^- \rightarrow 0$. One of the perturbative solutions becomes the minimal area solution for $\Delta x^+ = 20$ at some finite $\Delta x^-$. In every case, the minimal area plateaus to a finite, non-zero value as $\Delta x^- \rightarrow 0$.

\section{Toy Model with $\Delta K=\Delta S \neq 0$}
\label{toyy}

In this appendix, we present a toy model with a countable number of degrees of freedom (qubits), which shares the property we found for interacting theories on a null slab: $\Delta K = \Delta S \ne 0$. This relation is only possible as a limiting statement, because zero relative entropy between two states $\rho^1$ and $\rho^0$ implies that the states are equal. Moreover, the relation requires an infinite number of degrees of freedom, or else it would be reached before the limit is taken, in contradiction with the previous sentence. 

To demonstrate the effect in a toy model, we construct a decreasing sequence of algebras $A_n$ from which any fixed bounded operator will disappear as $n\rightarrow \infty$; in other words, $\cap A_n$ only contains multiples of the identity operator. Consider an infinite sequence of qubits. The algebra generated by the qubits operators for the qubits at position $n,n+1,...$ will be denoted by $A_n$. The algebras are nested: $A_m \subset A_n$ for $m>n$. The relative entropy of two states reduced to $A_n$ will decrease with $n$; that is, $\Delta K - \Delta S \rightarrow 0$. 

Consider states which are formed by tensor products of two-qubit states for the $k^{\text{th}}$ and $(k^2)^{\text{th}}$ qubits. This choice for the entanglement is arbitrary, but entanglement between the qubits $k$ and $f(k)$ with $f(k)$ growing much faster than $k$ is necessary to generate more entanglement than the entropy that is lost as we trace over the first $n$ qubits. 

Entanglement plays an important role in keeping $\Delta S$ finite while the relative entropy goes to zero.  The classical entropy is monotonous, so without quantum entanglement, the entropies must tend to zero with increasing $n$. In the quantum case, the entropy is no longer monotonous, but the relative entropy is monotonous and tends to zero instead. 

Consider generic states of the form
\be
\rho=\bigotimes_{\substack{ i \\ i \not = k^2}} \rho_{i,i^2}\,\label{formrho}.
\ee  
In this tensor product, we omit $i$ if $i$ is already included in the product by a previous factor of $\rho_{k,i}$ with $k^2=i$. The global relative entropy of two states both of the form in Eq.~\eqref{formrho} is 
\be \label{pairbtwo}
S(\rho^1|\rho^0)=\sum_i S(\rho^1_{i,i^2}|\rho^0_{i,i^2})\,.
\ee
We want a finite relative entropy, so a convergent series. We construct a sequence of mixed states $\rho_n$ by tracing over the first $n-1$ qubits of $\rho$. As $n$ tends to infinity, the relative entropy approaches zero:
\be
0 \le S(\rho^1_n|\rho^0_n)<\sum_{k=\sqrt{n}}^\infty S(\rho^1_{k,k^2}|\rho^0_{k,k^2})\rightarrow 0\label{pairsbtwo}\,,
\ee
where we have used that the positivity of the relative entropy, and the fact that the relative entropy of two states on a pair of qubits is greater than that of the states reduced to the second qubit of the pair. 

We want the global $\Delta S$ to remain finite as $n$ goes to infinity:
\be \ 
\Delta S=\sum_k \Delta S(k,k^2)<\infty\,.\label{er}
\ee

For the sequence of entropies $\Delta S_n$, we have 
\be
\Delta S_n=\sum_{k=n}^\infty \Delta S(k,k^2)+\sum_{k=\sqrt{n}}^n \Delta S_{\text{red}}(k,k^2)\,.\label{sumi}
\ee
Here, $\Delta S_\text{red}$ denotes the entropy of the reduced states on the second qubit of the pairs. The pairs of qubits with $k<\sqrt{n}$ have been completely traced out, while the pairs with $k>n$ are still completely included in the state.

Using Eq.~\eqref{er}, we see that the first sum in Eq.~\eqref{sumi} tends to zero as $n\rightarrow \infty$. For the second sum to have a finite and positive limit, we demand
\be
\Delta S_{\textrm{red}}(k,k^2)\sim \frac{c}{k \log k}\,,\label{rate}
\ee
which gives 
\be
\sum_{k=\sqrt{n}}^n \Delta S_{\textrm{red}}(k,k^2)\sim c(\log\log n-\log\log \sqrt{n})=c \log 2\,.
\ee
If $\Delta S_{\textrm{red}}(k,k^2)$ decays much faster, we get $\lim \Delta S_n=0$, which is not what we want. To get a non-zero answer, the entropy of the pairs $\Delta S_{\textrm{red}}(k,k^2)$ must not be integrable. If it decays at a slower asymptotic rate than Eq.~\eqref{rate}, the limiting value of $\Delta S_n$ is infinity. 

Now we choose the two qubit states $\rho^{0}_{k,k^2}$ and $\rho^{1}_{k,k^2}$. We impose three conditions: the relative entropies of these pairs should be integrable,
Eq. \eqref{pairsbtwo}; the differential entropies $\Delta S$ of these pairs should also be integrable, Eq.~\eqref{er}; and the $\Delta S_{\textrm{red}}$ of the states on the second qubit should have the asymptotic form in Eq.~\eqref{rate}, or slower than this if we want to obtain $\lim \Delta S_n \rightarrow \infty$.   

We choose the pair of states to be 
\bea
\rho^0=p |\psi\rangle\langle\psi|+(1-p) |\phi\rangle\langle\phi|\,,\\
\rho^1=p^\prime |\psi\rangle\langle\psi|+(1-p^\prime) |\phi\rangle\langle\phi|\,,
\eea
with $|\phi\rangle$, $|\psi\rangle$ a pair of orthogonal pure states for the two qubits. (We choose mixed states because the relative entropy diverges for any two non-identical pure states.) Taking $\delta p\equiv p^\prime-p$ to be small, we find
\begin{eqnarray}
	S(\rho^1|\rho^0)\simeq \frac{\delta p^2}{2 p (1-p)}\label{uno}\,,\\
	\Delta S\simeq\delta p \, \log\left(\frac{1-p}{p}\right)\label{dos}\,.
\end{eqnarray}
Here, $p$ and $\delta p$ depend on the pair of qubits $(k,k^2)$, and the dependence on $k$ is such that these entropies are integrable.

To evaluate the reduced entropy, we have to specify the pure states in terms of the qubits. The choice is arbitrary, but there are some restrictions. We cannot choose two orthogonal maximally entangled states for $|\psi\rangle$ and $|\phi\rangle$, because in this case, the reduced density matrices $\rho^0$ and $\rho^1$ will both equal $\tfrac{1}{2}{\bf I}$ and we obtain $\Delta S_{\textrm{red}} = 0$.  Instead, we take the two orthogonal states
\bea
|\psi\rangle&=&a |00\rangle+\sqrt{1-a^2} |11\rangle\,, \\
|\phi\rangle&=&b |01\rangle+\sqrt{1-b^2} |10\rangle\,.
\eea
Then the entropy is 
\be
\Delta S_{\textrm{red}}\simeq 2 (a-b)(a+b)\,\textrm{arctanh} (1-2 b^2 (1-p)-2 a^2 p) \,\,\delta p\,.
\ee

We can tune the dependence of $p,\delta p, a,b$ on $k$ so that the entropy goes as Eq.~\eqref{rate} and both Eq.~\eqref{uno} and Eq.~\eqref{dos} are integrable. We fix $a$ and $b$ and take $\delta p\simeq 1/(k\log k)$. The relative entropy is integrable because it contains a higher power of $\delta p$. For the total $\Delta S$ to be finite, we can choose $p\simeq 1/2+1/k$, to get an additional power of $1/k$ from the logarithm term in Eq.~\eqref{dos}. Then the states converge to a random state in the sub-Hilbert space spanned by $\left\{|\phi\rangle ,|\psi\rangle\right\}$. 
With this choice, both the total $\Delta S$ and the relative entropy are finite; the relative entropy goes to zero with $n$, while the limit of $\Delta S$ is
\be
\lim_{n \rightarrow \infty} \Delta S_n\rightarrow 2 (a-b)(a+b)\,\textrm{arctanh} (1- b^2 - a^2) \log(2)\,.
\ee
It is also clear that we have $\Delta K_n - \Delta S_n \rightarrow 0$ in the limit, or equivalently the relative entropy goes to zero.

The limit for $\Delta S_n$ can be made much larger (or infinite) by slowing the asymptotic decay of $\delta p$.  For example, keeping $a$, $b$ and $p$ as before but setting 
 $\delta p=1/k$ causes $\Delta S_n$ to diverge with finite initial $\Delta S$, while the relative entropy remains constant.

\bibliographystyle{utcaps}
\bibliography{all}

\end{document}